\documentclass[12pt]{article}
\usepackage{graphicx}
\usepackage{amssymb}
\usepackage{amsmath}

\begin{document}

\newcommand{\bear}{\begin{array}}  
\newcommand {\eear}{\end{array}}
\newcommand{\bea}{\begin{eqnarray}}   
\newcommand{\eea}{\end{eqnarray}}
\newcommand{\beq}{\begin{equation}}   
\newcommand{\eeq}{\end{equation}}
\newcommand{\bef}{\begin{figure}}  \newcommand 
{\eef}{\end{figure}}
\newcommand{\bec}{\begin{center}}  \newcommand 
{\eec}{\end{center}}
\newcommand{\non}{\nonumber}  \newcommand 
{\eqn}[1]{\beq {#1}\eeq}
\newcommand{\la}{\left\langle}  
\newcommand{\ra}{\right\rangle}
\newcommand{\ds}{\displaystyle}

\def\SEC#1{Sec.~\ref{#1}}
\def\FIG#1{Fig.~\ref{#1}}
\def\EQ#1{Eq.~(\ref{#1})}
\def\EQS#1{Eqs.~(\ref{#1})}
\def\lrf#1#2{ \left(\frac{#1}{#2}\right)}
\def\lrfp#1#2#3{ \left(\frac{#1}{#2} 
\right)^{#3}}
\def\GEV#1{10^{#1}{\rm\,GeV}}
\def\MEV#1{10^{#1}{\rm\,MeV}}
\def\KEV#1{10^{#1}{\rm\,keV}}



\vskip 1.35cm

\begin{center}
{\large \bf
A note on
``The Affleck-Dine Mechanism in Conformally Sequestered Supersymmetry Breaking''
 }
\vskip 1.2cm

Kazunori Nakayama$^a$
and
Fuminobu Takahashi$^b$

\vskip 0.4cm

{ \it $^a$Institute for Cosmic Ray Research,
University of Tokyo, Kashiwa 277-8582, Japan}\\
{\it $^b$Institute for the Physics and Mathematics of the Universe,
University of Tokyo, Kashiwa 277-8568, Japan}
\date{\today}


\end{center}


In Ref.~\cite{Nakayama:2009ar} we proposed a mechanism to create baryon asymmetry 
in a conformally sequestered supersymmetry (SUSY) breaking scenario
through the Affleck-Dine (AD) mechanism.
After the submission, however, we found a critical error in our discussion.

In our scenario, the AD mass depends on the energy scale at which the operator
like Eq.~(6) of the original paper is evaluated.  
In Eq.~(7) we assumed that the energy scale corresponds to the AD field value
and this assumption made it possible to avoid the trapping of the AD field
by the color breaking minimum,
which was a problem of the original AD mechanism in anomaly-mediated SUSY breaking.
This assumption, however, was not valid,
since the typical energy scale of the operator like Eq.~(6) should be that of the
{\it hidden} sector, which is not related to the AD field value.
Thus Eq.~(8) does not hold and our solution to the problem of the color breaking minimum
in the AD mechanism does not work.

We thank the anonymous referee who noted a critical point described above.

{}

\end{document}